\documentclass[prc,superscriptaddress,twocolumn,showpacs,preprintnumbers]{revtex4}

\usepackage[dvips]{graphicx}
\usepackage{amsmath}
\usepackage{amssymb}
\usepackage{times}
\usepackage{mathrsfs}
\usepackage{multirow}
\usepackage{bm}

\newcommand{\beq}{\begin{equation}}
\newcommand{\eeq}{\end{equation}}
\newcommand{\bea}{\begin{eqnarray}}
\newcommand{\eea}{\end{eqnarray}}

\def\mbold#1{\mbox{\boldmath $#1$}}

\def\beq{\begin{equation}}
\def\eeq{\end{equation}}
\def\bea{\begin{eqnarray}}
\def\eea{\end{eqnarray}}
\def\fun#1#2{\lower3.6pt\vbox{\baselineskip0pt\lineskip.9pt
  \ialign{$\mathsurround=0pt#1\hfil##\hfil$\crcr#2\crcr\sim\crcr}}}

\def\a{\alpha}

\def\mbold#1{\mbox{\boldmath $#1$}}

\begin{document}


\title{Effects of four-body breakup 
on $^6$Li elastic scattering near the Coulomb barrier
}

\author{Shin Watanabe}
\email[]{s-watanabe@phys.kyushu-u.ac.jp}
\affiliation{Department of Physics, Kyushu University, Fukuoka 812-8581, Japan}

\author{Takuma Matsumoto}
\email[]{matsumoto@phys.kyushu-u.ac.jp}
\affiliation{Department of Physics, Kyushu University, Fukuoka 812-8581, Japan}

\author{Kosho Minomo}
\email[]{minomo@phys.kyushu-u.ac.jp}
\affiliation{Department of Physics, Kyushu University, Fukuoka 812-8581, Japan}

\author{Masanobu Yahiro}
\email[]{yahiro@phys.kyushu-u.ac.jp}
\affiliation{Department of Physics, Kyushu University, Fukuoka 812-8581, Japan}

\date{\today}

\begin{abstract}
We investigate projectile breakup effects on $^6$Li+$^{209}$Bi 
elastic scattering near the Coulomb barrier with the four-body version of 
the continuum-discretized coupled-channels method (four-body CDCC). 
This is the first application of four-body CDCC to 
$^6$Li elastic scattering. 
The elastic scattering is well described 
by the $p$+$n$+$^4$He+$^{209}$Bi four-body model. 
We propose a reasonable three-body model for describing 
the four-body scattering, clarifying four-body dynamics 
of the elastic scattering. 
\end{abstract}


\maketitle


{\it Introduction.} 
Plenty of nuclei are considered to have two-cluster or 
three-cluster configurations as their main components. 
Three-cluster dynamics is, however, nontrivial 
compared with two-cluster dynamics. 
Systematic understanding of three-cluster dynamics is hence important. 
There are many nuclei that can be described by three-cluster models. 
For example, low-lying states of $^{6}$He and $^{6}$Li are explained by 
$N$ + $N$ + $^{4}$He three-body
models~\cite{Fun94,Hiy95,Myo01,Cso94,Ara95,Kik12}, where $N$ stands for
a nucleon. The comparison of
the two nuclei is important  to see the difference between 
dineutron 
and proton-neutron 
correlations. 
Two-neutron halo nuclei such as 
$^{11}$Li, $^{14}$Be, and $^{22}$C 
are reasonably described 
by an $n$ + $n$ + $X$ three-cluster model, where $X$ is a core nucleus. 
Properties of these three-cluster configurations 
should be confirmed by measuring scattering of the nuclei 
and analyzing the measured cross sections with accurate reaction theories. 
The reactions are essentially four-body scattering composed of three 
constituents of projectile and a target nucleus. 
Accurate theoretical description of four-body scattering 
is thus an important subject in nuclear physics.

The continuum-discretized coupled-channels method
(CDCC) is a fully quantum-mechanical method 
of describing not only three-body scattering 
but also four-body scattering~\cite{Kam86,Aus87,Yah12}. 
CDCC has succeeded in 
reproducing experimental data on 
both three- and four-body scattering. 
The theoretical foundation of CDCC is shown with 
the distorted Faddeev equation~\cite{Aus89,Aus96,Del07}. 
CDCC for four-body (three-body) scattering is often called 
four-body (three-body) CDCC; see 
Refs~\cite{Mor01,Mat03,Mat04,Ega04,Rod05,Mor06,Mat06,Rod08,Mor09,Rod09,Ega09,Mat09,Mat10} and references therein for four-body CDCC.  
So far four-body CDCC was applied to only $^6$He scattering.

For $^6$He + $^{209}$Bi scattering at 19 and 22.5~MeV near the Coulomb barrier, 
the measured total reaction cross sections are largely enhanced 
in comparison with that for $^6$Li + $^{209}$Bi 
scattering at 29.9 and 32.8~MeV 
near the Coulomb barrier~\cite{Agu00,Agu01}. 
Keeley \textit{et al.}~\cite{Kee03} analyzed the $^6$He~+~$^{209}$Bi scattering
with three-body CDCC in which the $^6$He~+~$^{209}$Bi system was 
assumed to be a $^2n$~+~$^{4}$He~+~$^{209}$Bi three-body system, i.e., 
a pair of extra neutrons in $^6$He was treated as a single particle, 
dineutron ($^2n$). 
The enhancement of the total reaction cross section of the
$^6$He~+~$^{209}$Bi scattering is  found to be due to the electric
dipole ($E1$) excitation of $^6$He to its continuum 
states~\cite{Rus05}, i.e., Coulomb breakup of $^6$He, which is almost absent 
in the $^6$Li~+~$^{209}$Bi scattering.
The three-body CDCC calculation, however, does not 
reproduce the angular distribution 
of the measured elastic cross section and overestimates the measured
total reaction cross section by a factor of 2.5. 
This problem is solved by four-body CDCC~\cite{Mat06} 
in which the total system is assumed to be an 
$n$~+~$n$~+~$^{4}$He~+~$^{209}$Bi four-body system.

The $^6$Li~+~$^{209}$Bi scattering near the Coulomb barrier was, meanwhile,  
analyzed with three-body CDCC by assuming a $d$~+~$^4$He~+~$^{209}$Bi 
three-body model~\cite{Kee03}.  
The three-body CDCC calculation could not reproduce the data
without normalization factors for the potentials between $^6$Li and
$^{209}$ Bi. 
This result indicates that four-body CDCC should be applied to 
the $^6$Li~+~$^{209}$Bi scattering.

In this paper, we analyze $^6$Li~+~$^{209}$Bi elastic scattering at
29.9 and 32.8~MeV with four-body CDCC by assuming 
the $p$~+~$n$~+~$^4$He~+~$^{209}$Bi four-body model. 
This is the first application of 
four-body CDCC to $^6$Li scattering. 
The four-body CDCC calculation reproduces 
the measured elastic cross sections, whereas 
the previous three-body CDCC calculation does not. 
Four-body dynamics of the elastic scattering is investigated, 
and it is discussed what causes the failure of 
the previous three-body CDCC calculation. 
Finally we propose a reasonable three-body model for describing the four-body 
scattering.

\begin{figure}[htbp]
\includegraphics[width=0.4\textwidth,clip]{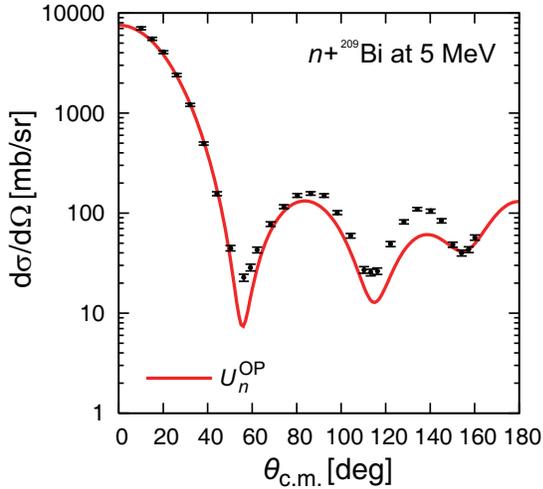}
\caption{Angular distribution of elastic cross section 
for $n$ + $^{209}$Bi scattering at 5 MeV. 
The solid line is the result of the neutron optical potential 
$U_n^{\rm OP}$. 
The experimental data is taken from Ref.~\cite{Ann85}. 
}
\label{n-209Bi}
\end{figure}

{\it Theoretical framework.}
One of the most natural frameworks to describe $^6$Li~+~$^{209}$Bi scattering 
is the $p$ ~+~ $n$ + $^4$He + $^{209}$Bi four-body model. 
Dynamics of the scattering 
is governed by the Schr\"{o}dinger equation
\begin{eqnarray}
(H-E)\Psi=0
\label{original-H}
\end{eqnarray}
for the total wave function $\Psi$, where $E$ is a total energy of the system. 
The total Hamiltonian $H$ is  defined by 
\bea
 H=K_{R}+U+h 
\label{H4}
\eea
with 
\bea
U=U_{n}(R_{n})+U_{p}(R_{p})+ U_{\alpha}(R_{\alpha})
+\frac{e^2Z_{\rm Li}Z_{\rm Bi}}{R}, 
\label{potU}
\eea
where $h$ denotes the internal Hamiltonian of $^6$Li, 
${\mbold R}$ is the center-of-mass coordinate of $^{6}$Li
relative to $^{209}$Bi, $K_{R}$ stands for the kinetic energy 
operator associated with ${\mbold R}$,  
and $U_{x}$ describes the nuclear part of the optical potential
between $x$ and $^{209}$Bi as a function of the relative coordinate 
$R_{x}$. 
As $U_\alpha$, we adopt the optical potential of Barnett and 
Lilley~\cite{Bar74}. 
Parameters of $U_n$ are fitted to reproduce 
experimental data~\cite{Ann85} on $n$~+~$^{209}$Bi elastic scattering at 5~MeV, 
where only the central interaction is taken for simplicity. 
As shown in Fig.~\ref{n-209Bi}, the neutron optical potential 
$U_n^{\rm OP}$ thus fitted is consistent with the data. 
The resultant parameter set is the same as that in the global optical
potential of Koning and Delaroche~\cite{Kon03}, 
except parameters $a_{V}$, $W_V$ and $W_D$ are changed into 0.55~fm, 
0~MeV and 4.0~MeV, respectively.  
The proton optical potential $U_p$ is assumed to be the same as $U_n$. 
Coulomb interactions in the $p$-$^{209}$Bi and $\alpha$-$^{209}$Bi 
subsystems are approximated 
into $e^2Z_{\rm Li}Z_{\rm Bi}/R$, because Coulomb breakup effects are 
negligibly small in the present scattering.

The internal Hamiltonian $h$ of $^6$Li is described by the $p$ + $n$ + $^4$He 
orthogonality condition model~\cite{Sai69}. 
The Hamiltonian of $^6$Li agrees with 
that of $^6$He in Ref.~\cite{Mat06}, when 
the Coulomb interaction between $p$ and $^4$He is neglected. 
Namely, the Bonn-A interaction~\cite{Mac89} is taken in the $p$-$n$ 
subsystem and the so-called KKNN interaction~\cite{Kan79} is used 
in the $p$-$\alpha$ and $n$-$\alpha$ subsystems, 
where the KKNN interaction is 
determined from experimental data on low-energy
nucleon-$\alpha$ scattering.

Eigenstates of $h$ consist of finite number of discrete states 
with negative energies and continuum states with positive energies. 
In four-body CDCC, the continuum states of projectile 
are discretized into a finite number of pseudostates 
by either 
the pseudostate method~\cite{Mat03,Mat04,Ega04,Mat06,Ega09,Mat09,Mat10,Mor01,Rod05,Mor06,Rod08,Mor09} or the momentum-bin method~\cite{Rod09}. 
The Schr\"{o}dinger equation \eqref{original-H} is 
solved in a modelspace ${\cal P}$ spanned 
by the discrete and discretized-continuum states:
\begin{eqnarray}
{\cal P}(H-E){\cal P}\Psi_{\rm CDCC}=0 .
\label{H-CDCC}
\end{eqnarray} 
In the pseudostate method, the discrete and discretized continuum states 
are obtained by diagonalizing $h$ in a space spanned by 
$L^2$-type basis functions. As the basis function, 
the Gaussian~\cite{Mat03,Mat04,Ega04,Mat06,Ega09,Mat09,Mat10} or the transformed Harmonic Oscillator function~\cite{Mor01,Rod05,Mor06,Rod08,Mor09} is 
usually taken. 
In this paper, we use the Gaussian function. 
The modelspace ${\cal P}$ is then described by 
\begin{eqnarray}
 {\cal P}&=&\sum_{nIm}|\Phi_{nIm}\rangle\langle\Phi_{nIm}|,
\end{eqnarray}
where $\Phi_{nIm}$ is the $n$th eigenstate of $^6$Li with an energy 
$\epsilon_{nI}$, a total spin $I$ and its projection on the $z$-axis $m$.

In actual calculations, the $\Phi_{nIm}$ are obtained for 
$I^\pi=1^+$, $2^+$ and $3^+$ by diagonalizing $h$ with 10 Gaussian functions 
for each coordinate in which the range parameters are taken 
form 0.1 fm to 12 fm in geometric series. 
The $\Phi_{nIm}$ with $\epsilon_{nI} \leq 20$ MeV are excluded from 
${\cal P}$, since they do not affect cross sections of 
$^6$Li + $^{209}$Bi scattering. 
The resulting numbers of the discrete states are 
65 (including the ground state of $^6$Li), 57 and 63 
for $1^+$, $2^+$, and $3^+$ states, respectively.

The CDCC wave function $\Psi_{\rm CDCC}^{JM}$, 
with the total angular momentum $J$ and its projection on the $z$-axis $M$, 
are expressed as 
\begin{eqnarray}
 \Psi^{JM}=\sum_{nIL}
\chi_{nIL}^J (P_{nI},R)/R
\;{\cal Y}_{nIL}^{JM}
\end{eqnarray}
with 
\bea
{\cal Y}_{\gamma}^{JM}=[\Phi_{nI}({\mbold y}_c,
{\mbold r}_c)\otimes i^LY_L(\hat{\mbold R})]_{JM}
\eea
for the orbital angular momentum $L$ regarding ${\mbold R}$. 
The expansion-coefficient $\chi_{\gamma}^J$, where 
$\gamma=(n,I,L)$, describes a motion of $^6$Li 
in its $(n,I)$ state 
with linear and angular relative momenta $P_{nI}$ and $L$. 
Multiplying the four-body
Schr\"{o}dinger equation \eqref{H-CDCC} by
${\cal Y}_{\gamma'}^{*JM}$ from the left and integrating it over
all variables except $R$, one can obtain a set of coupled
differential equations for $\chi_{\gamma}^J$:
\begin{eqnarray}
\bigg[\frac{d^2}{dR^2}
-\frac{L(L+1)}{R^2}
-\frac{2\mu}{\hbar^2}U_{\gamma\gamma}(R)
+P_{nI}^2
\bigg]
\chi_{\gamma}^J(P_{nI},R)
\nonumber \\
&&\hspace{-55mm}
=
\frac{2\mu}{\hbar^2}\sum_{\gamma'\ne \gamma}
U_{\gamma'\gamma}(R)\chi_{\gamma'}^J(P_{n'I'},R)
\label{CDCCeq}
\end{eqnarray}
with the coupling potentials 
\begin{equation}
U_{\gamma'\gamma}(R)=
\langle {\cal Y}_{\gamma'}^{JM}|
U_{n}(R_{n})+U_{p}(R_{p})+U_{\alpha}(R_{\alpha})
|{\cal Y}_{\gamma}^{JM} \rangle,
\nonumber
\end{equation}
where $\mu$ is the reduced mass between $^6$He and $^{209}$Bi.
The elastic and discrete breakup $S$-matrix elements are obtained 
by solving Eq.~(\ref{CDCCeq}) under the standard asymptotic
boundary condition~\cite{Kam86,Piy89}.

We also do three-body CDCC calculations by assuming 
a $d$ + $^{4}$He + $^{209}$Bi model, following Refs.~\cite{Kee03,Rus05}. 
As an interaction between $d$ and
$^4$He, we take the potential of Ref.~\cite{Sak86}, 
which are determined from experimental data on 
the ground-state energy ($-1.47$ MeV) and the $3^+$-resonance state energy 
($0.71$ MeV) of $^6$Li and low-energy $d$-$\alpha$ scattering phase shifts.
The continuum states between $d$ and $^4$He are discretized 
with the pseudostate method~\cite{Mat03} and are truncated 
at 20 MeV in the excitation energy of $^6$Li from the $d$-$^4$He
threshold.
The $d$-$^{209}$Bi (type-a) optical potential ($U_d^{\rm OP}$)~\cite{Bud63} 
is taken as $U_d$, whereas 
$U_{\alpha}$ is common between three- and four-body CDCC 
calculations.

\begin{figure}[htbp]
\includegraphics[width=0.4\textwidth,clip]{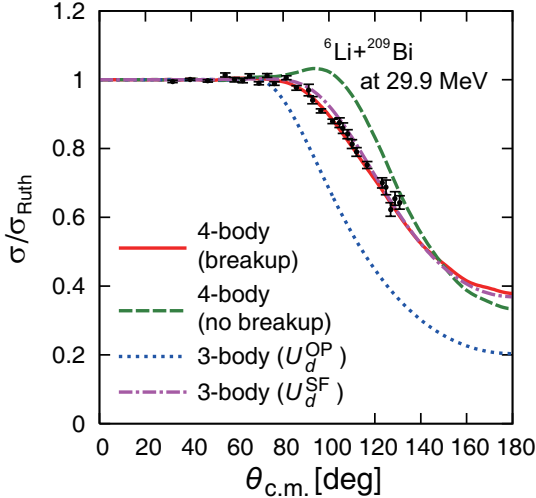}
\caption{Angular distribution of the elastic cross section
 for $^6$Li~+~$^{209}$Bi scattering at 29.9 MeV. 
 The cross section is normalized by the Rutherford cross section. 
 The dotted (dot-dashed) line stands for 
 the result of three-body CDCC calculation in which 
 $U_d^{\rm OP}$ ($U_d^{\rm SF}$) is taken as $U_d$. 
 The solid (dashed) lines represent the results of 
  four-body CDCC calculations with (without) breakup effects. 
  The experimental data are taken from Ref.~\cite{Agu00,Agu01}.
}
\label{elastic1}
\end{figure}

\begin{figure}[htbp]
\includegraphics[width=0.4\textwidth,clip]{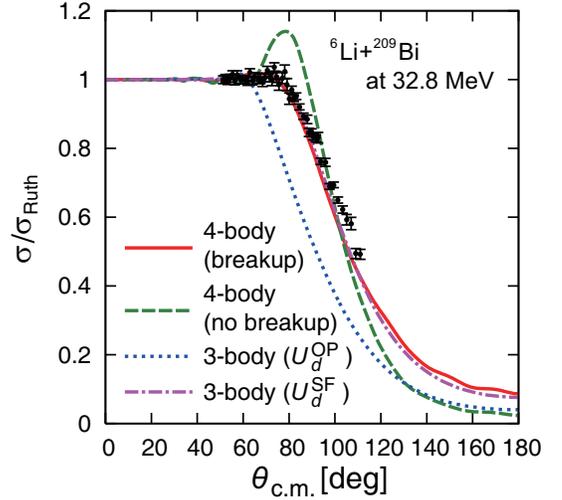}
\caption{The same as in Fig.~\ref{elastic1} but for $^6$Li~+~$^{209}$Bi
 scattering at 32.8 MeV. The experimental data are taken from
 Ref.~\cite{Agu00,Agu01}.
 }
\label{elastic2}
 \end{figure}

{\it Results.}
Figure~\ref{elastic1} shows the angular distribution of elastic 
cross section for $^6$Li~+~$^{209}$Bi scattering at 29.9~MeV. 
The dotted line shows the result of three-body CDCC calculation 
with $U_d^{\rm OP}$ as $U_d$. 
This result, which is consistent with the previous result 
of Ref.~\cite{Kee03}, 
underestimates the measured cross section~\cite{Agu00,Agu01}. 
The solid (dashed) line, meanwhile, stands for the result of 
four-body CDCC calculation with (without) projectile breakup effects. 
In CDCC calculations without $^6$Li-breakup, 
the modelspace ${\cal P}$ is composed only of the $^6$Li ground state. 
The solid line reproduces the experimental cross section, but 
the dashed line does not.
The projectile breakup effects are thus significant and the present 
$^6$Li scattering is well described by 
the $p$~+~$n$~+~$^4$He~+~$^{209}$Bi four-body model. 
This conclusion is true also for $^6$Li~+~$^{209}$Bi scattering 
at 32.8~MeV shown in Fig.~\ref{elastic2}.

 \begin{figure}[htbp]
\includegraphics[width=0.4\textwidth,clip]{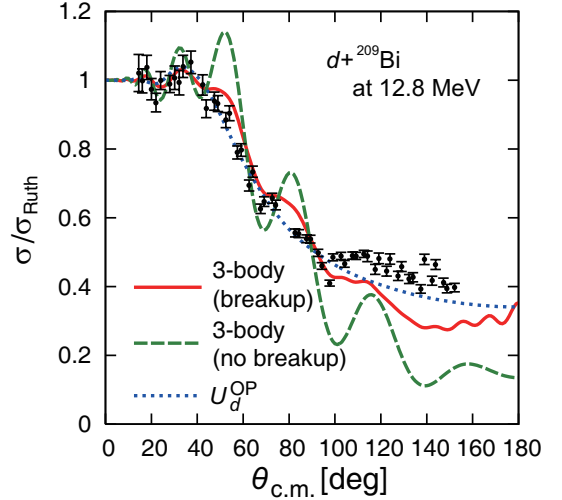}
\caption{Angular distribution of the elastic cross section
 for $d$~+~$^{209}$Bi scattering at 12.8~MeV. 
 The solid (dashed) line stands for the result of 
 three-body CDCC calculation with (without) deuteron breakup, whereas 
 the dotted line is the result of the deuteron optical potential 
 $U_d^{\rm OP}$. 
 The experimental data are taken from Ref.~\cite{Bud63}.
 }
\label{elastic-d}
 \end{figure}

Now we consider $d$-breakup in the $^6$Li scattering in 
order to understand four-body dynamics of the scattering. 
In the limit of no $d$-breakup, the interaction between $d$ and 
$^{209}$Bi can be obtained by folding $U_n$ and $U_p$ 
with the deuteron density. 
This potential is referred to as the single-folding potential $U_d^{\rm SF}$. 
In Figs.~\ref{elastic1} and \ref{elastic2}, the dot-dashed lines show 
the results of three-body CDCC calculations with $U_d^{\rm SF}$ as $U_d$. 
The results well simulate those of four-body CDCC calculations, 
i.e., the solid lines. This indicates that $d$-breakup is 
suppressed in the $^6$Li scattering. 
Intuitive understanding of this property is as follows. 
As a characteristic of the present $^6$Li scattering, it is quite peripheral 
in virtue of the Coulomb barrier. 
The scattering is dominated by the configuration in which 
$\a$ is located between $d$ and the target, 
because $U_{\a}$ is more attractive than $U_{d}$. 
In this configuration, $d$ is out of the range of $U_{n}$ and $U_{p}$, 
so that $d$-breakup is suppressed. 
The $^6$Li elastic scattering near the Coulomb barrier is thus well 
described by the $d$~+~$\a$~+~$^{209}$Bi three-body model, 
if $U_d^{\rm SF}$ is taken as $U_d$.

Figure \ref{elastic-d} shows the angular distribution of 
elastic cross section for $d$~+~$^{209}$Bi scattering at 12.8~MeV. 
The solid and dashed lines stand for the results 
of three-body CDCC calculations 
with and without $d$-breakup, respectively, 
in which the $p$~+~$n$~+~$^{209}$Bi model is assumed and 
both Coulomb and nuclear breakup are taken into account. 
The solid line reproduces the data fairly well, 
but the dashed line does not.
Thus $d$-breakup is significant for the deuteron scattering. 
The deuteron optical potential $U_d^{\rm OP}$ (dotted line) 
yields fairly good agreement with the data, but the radius of 
$U_d^{\rm OP}$ is larger than that of $U_d^{\rm SF}$. This is the reason 
why three-body CDCC calculations with $U_d^{\rm OP}$ as $U_d$ 
can not reproduce the measured elastic cross section 
for $^6$Li~+~$^{209}$Bi scattering. 
The difference between $U_d^{\rm SF}$ and $U_d^{\rm OP}$ mainly comes from 
the fact that $U_d^{\rm OP}$ includes $d$-breakup effects, 
whereas $U_d^{\rm SF}$ does not.

{\it Summary.} 
The $^6$Li~+~$^{209}$Bi scattering at 29.9 MeV and 32.8 MeV 
near the Coulomb barrier are well described by 
four-body CDCC based on the $p$~+~$n$~+~$^4$He~+~$^{209}$Bi model. 
This is the first application of four-body CDCC to $^6$Li scattering. 
In the $^6$Li scattering, $d$-breakup is strongly 
suppressed, suggesting that the $d$~+~$^4$He~+~$^{209}$Bi model becomes 
good, if the single-folding potential $U_d^{\rm SF}$ with no 
$d$-breakup is taken as an interaction between $d$ and the target. 
For $d$+~$^{209}$Bi scattering at 12.8~MeV, meanwhile, 
$d$-breakup is significant, so that the deuteron optical potential 
$U_d^{\rm OP}$ includes $d$-breakup effects. 

Four-body CDCC is applicable also for 
$n$ + $^6$Li scattering that is a key reaction 
in nuclear engineering. In the scattering, 
$^6$Li breakup into $n$ + $p$ + $\alpha$ is considered to be not negligible 
for emitted neutron spectra~\cite{Mat11}. 
We will discuss this point in a forthcoming paper.


The authors would like to thank Y. Watanabe, K. Ogata and K. Kat\=o 
for helpful discussions. 
This work has been supported in part by the Grants-in-Aid for
Scientific Research of Monbukagakusyou of Japan and JSPS.



\end{document}